\documentclass[epj,nopacs]{svjour}
\pdfoutput=1

\usepackage{times}
\usepackage[caption=false]{subfig}
\usepackage{graphicx}
\usepackage{hyphenat}
\usepackage{xspace}
\usepackage{textcomp}
\usepackage[numbers,compress]{natbib}
\usepackage{rotating}
\usepackage[colorlinks,breaklinks]{hyperref}

\newcommand{\antikt}{anti-$k_t$\xspace}
\newcommand{\pt}{\ensuremath{p_{\rm T}}\xspace}
\newcommand{\y}{\ensuremath{y}\xspace}
\newcommand{\mjj}{\ensuremath{m_{\rm jj}}\xspace}
\newcommand{\ptrel}{\ensuremath{\pt^{\rm rel}}\xspace}
\newcommand{\bbar}{\ensuremath{\overline{b}}\xspace}
\newcommand{\bbbar}{\ensuremath{b\overline{b}}\xspace}
\newcommand{\ttbar}{\ensuremath{t\overline{t}}\xspace}
\newcommand{\pythia}{Pythia\xspace}
\newcommand{\herwig}{Herwig\xspace}
\newcommand{\powheg}{POW\-HEG\xspace}
\newcommand{\mcatnlo}{MC\-@NLO\xspace}
\newcommand{\powhegpythia}{\powheg + \pythia}
\newcommand{\powhegherwig}{\powheg + \herwig}
\newcommand{\mcatnloherwig}{\mcatnlo + \herwig}

\title{\vspace{-3.0cm}\flushleft{\normalsize\normalfont{CERN-PH-EP-2011-146}} \flushright{\vspace{-0.86cm}\normalsize\normalfont{Submitted to Eur. Phys. J. C}} \\[3.0cm] \flushleft{Measurement of the inclusive and dijet cross-sections of $b$-jets in $pp$ collisions at $\sqrt{s}=7$\,TeV with the ATLAS detector}}
\titlerunning{Inclusive and dijet cross-sections of $b$-jets}

\author{The ATLAS Collaboration}
\institute{}
\date{December 19, 2011}

\abstract{
The inclusive and dijet production cross-sections have been measured for jets containing $b$-hadrons ($b$-jets) in 
proton--proton collisions at a centre-of-mass energy of $\sqrt{s} = 7$\,TeV, using the ATLAS detector at the LHC. The 
measurements use data corresponding to an integrated luminosity of $34$\,pb$^{-1}$. The $b$-jets are identified using either 
a lifetime-based method, where secondary decay vertices of $b$-hadrons in jets are reconstructed using information from the 
tracking detectors, or a muon-based method where the presence of a muon is used to identify semileptonic decays of 
$b$-hadrons inside jets. The inclusive $b$-jet cross-section is measured as a function of transverse momentum in the range 
$20 < \pt < 400$\,GeV and rapidity in the range $|y| < 2.1$. The \bbbar-dijet cross-section is measured as a function of 
the dijet invariant mass in the range $110 < \mjj < 760$\,GeV, the azimuthal angle difference between the two jets and the 
angular variable $\chi$ in two dijet mass regions. The results are compared with next-to-leading-order QCD predictions. Good 
agreement is observed between the measured cross-sections and the predictions obtained using \powhegpythia.
\mcatnloherwig shows good agreement with the measured \bbbar-dijet cross-section. However, it does not reproduce the 
measured inclusive cross-section well, particularly for central $b$-jets with large transverse momenta.
}

\begin{document}

\maketitle


\section{Introduction}

The production of $b$-quarks in proton--proton collisions at the Large Hadron Collider (LHC) provides an important test of 
perturbative QCD (pQCD). Calculations of the $b$-quark production cross-section have been performed at next-to-leading 
order (NLO) in pQCD~\cite{Frixione:1997ma}. These calculations can be matched to different parton\hyp{}shower and 
hadronisation models to produce final states that can be compared to those measured in collision data. 

Cross-sections for $b$-jet production in high energy $p\overline p$ collisions have been measured at the 
${\rm Sp\overline pS}$~\cite{Albajar:1986iu,Albajar:1988th} and 
Tevatron~\cite{Abe:1993hr,Abe:1993sj,Abbott:1999se,Abbott:2000iv} colliders. The experiments measured 
cross-sections different from those predicted by QCD at the time. This led to substantial improvements in the experimental 
methods and theoretical calculations. It is therefore of great interest to test the theoretical predictions at the higher 
centre-of-mass energy provided by the LHC. Moreover, the measurement of the $b$-jet cross-sections is an important 
ingredient in understanding other processes involving the production of $b$-quarks, which represent substantial backgrounds in 
many searches for new physics. Measurements of $b$-hadron production at $\sqrt{s} = 7$\,TeV in the forward region have 
been reported by LHCb~\cite{Aaij:2010gn} and in the central region by CMS~\cite{Khachatryan:2011hf,Khachatryan:2011wq}.

This paper describes measurements of the inclusive $b$-jet and $b\overline b$-dijet production cross-sections performed with
the ATLAS detector at the LHC. Jets are reconstructed from energy clusters in the calorimeter using the \antikt 
algorithm~\cite{Cacciari:2008gp}, with jet radius parameter $R=0.4$. The relatively long lifetime of hadrons containing 
$b$-quarks is exploited to obtain a jet sample enriched in $b$-jets by selecting jets with a reconstructed secondary vertex 
significantly displaced from the primary vertex. The number of $b$-jets in this enriched sample is derived from a fit to the 
invariant mass distribution of the charged particle tracks in the secondary vertex, assuming the pion mass for the individual 
particles. This is referred to as secondary vertex mass hereafter.

The inclusive cross-section is measured for jets containing $b$- or $\bbar$-quarks as a function of the transverse momentum, 
\pt, and rapidity, \y, for jets with $20 < \pt < 400$\,GeV and $|y| < 2.1$. The requirement $|y| < 2.1$ ensures that jets are 
contained within the acceptance of the inner tracking detectors. In the kinematic region $30 < \pt < 140$\,GeV, 
muon\hyp{}based $b$-tagging is used to provide a complementary, and largely independent, cross-section measurement as a 
function of jet \pt.

The \bbbar-dijet cross-section is measured for the leading and sub-leading jet in the event as a function of the dijet 
invariant mass, \mjj, the azimuthal angle difference between the two jets, $\Delta\phi$, and the angular variable 
$\chi = \exp{\left|y_1 - y_2\right|}$ for jets with $\pt > 40$\,GeV and $|y| < 2.1$. The variable $\chi$ is defined such that the 
cross-section of $2\to 2$ elastic scattering of point-like massless particles is approximately constant as a function of 
$\chi \simeq \frac{1 + \cos\theta^*}{1 - \cos\theta^*}$, where $\theta^*$ is the centre-of-mass scattering angle. To 
measure the cross-sections as a function of $\chi$, an additional acceptance requirement is used that restricts the boost of 
the dijet system to $|y_{\rm boost}| = \frac{1}{2}|y_1 + y_2| < 1.1$. This reduces the sensitivity to parton distribution 
function (PDF) uncertainties at small values of $x$, where $x$ is the fraction of the proton's momentum carried by the parton 
participating in the hard scattering. The resulting angular distributions provide a test of pQCD that is relatively insensitive to 
PDF uncertainties.

The measured cross-sections are corrected for all experimental effects using simulated events, to allow comparison with 
theoretical predictions. 

The data used for these measurements were collected by the ATLAS detector in 2010 and correspond to an integrated 
luminosity of $34.0 \pm 1.2$\,pb$^{-1}$. A detailed description of the luminosity determination can be found in 
Refs.~\cite{Aad:2011dr,ATLAS-CONF-2011-011}.


\section{The ATLAS detector}
The ATLAS detector~\cite{Aad:2008zzm} consists of an inner tracking system, immersed in a 2\,T axial magnetic field, 
surrounded by electromagnetic calorimeters, hadronic calorimeters and a muon spectrometer. The ATLAS reference system 
has the origin at the nominal interaction point. The $x$- and $y$-axes define the transverse plane, the azimuthal angle $\phi$ 
is measured around the beam axis, $z$, and the polar angle $\theta$ with respect to the $z$-axis. The pseudorapidity is 
defined as $\eta = -\ln\left(\tan\left(\theta/2\right)\right)$.

The inner detector (ID) has full coverage in $\phi$ and covers the pseudorapidity range $|\eta| < 2.5$. The ID consists of 
silicon pixel and microstrip detectors, surrounded by a transition radiation tracker (up to $|\eta| = 2.0$). The electromagnetic 
calorimeter is a lead-liquid argon sampling calorimeter covering $|\eta| < 3.2$. Hadronic calorimetry in the barrel 
($|\eta| < 1.7$) is provided by a scintillator tile calorimeter using steel as the absorber material. The end-cap hadronic 
calorimeter uses liquid argon with copper absorber plates and extends up to $|\eta| = 3.2$. Additional forward calorimeters 
extend the calorimetric coverage to $|\eta| < 4.9$, outside the acceptance of this measurement. The outer region of the 
detector is formed by a muon spectrometer that uses a toroidal magnetic field with a bending power of 1.5--5.5\,Tm in the barrel 
and 1.0--7.5\,Tm in the end-caps. Three layers of muon chambers provide precision tracking in the bending plane up to 
$|\eta| = 2.7$ and the trigger for muons up to $|\eta| = 2.4$.

The trigger system uses three consecutive trigger levels to record a selection of interesting events. The first level trigger (L1) is 
based on custom\hyp{}built hardware that processes the data with a fixed latency of $2.5$\,\textmu s. The second level and the 
event filter, collectively referred to as the high level trigger (HLT), are software-based triggers running on computing farms. Their 
average execution times are $40$\,ms and $4$\,s respectively, with a design output rate of $3$\,kHz and $200$\,Hz respectively.

Most of the events used in the measurements presented here are selected by the calorimeter\hyp{}based triggers. At L1, the 
electromagnetic and hadronic calorimeters are read out using trigger towers with a granularity of 
$\Delta\phi \times \Delta\eta = 0.1 \times 0.1$, with jet identification based on transverse energy in a sliding window of 
$4\times 4$ or $8\times 8$ trigger towers. At the beginning of data-taking in 2010 only the L1 triggers were active, while in the 
later runs the HLT was used to refine the jet selection further. Events containing jets with $20 < \pt < 40$\,GeV were triggered 
using the minimum bias trigger scintillators (MBTS)~\cite{Aad:2010ir}. The MBTS consist of 32 scintillator counters arranged 
in two discs located at $\pm 3.56$\,m from the interaction point, covering $2.09 < |\eta| < 3.84$. The hit multiplicity in the 
MBTS provides a high\hyp{}efficiency trigger for jet events, independent of the jet \pt, with negligible bias.


\section{Monte Carlo samples and theoretical predictions}

Simulated events produced by the \pythia~6.423~\cite{Sjostrand:2006za} event generator are used for the baseline 
comparisons and to evaluate corrections. \pythia implements leading-order (LO) pQCD matrix elements for $2 \to 2$ 
processes, \pt-ordered parton\hyp{}showers calculated in a leading-logarithmic approximation and an underlying event 
simulation using multi-parton interactions. It uses the Lund string model~\cite{Andersson:1983ia} for hadronisation. All events 
were generated using a specially tuned set of parameters denoted as AMBT1~\cite{Aad:2010ir} with MRST 
LO$^{*}$~\cite{Sherstnev:2007nd} parton-density functions. The generated particles are passed through a full 
simulation~\cite{Aad:2010wqa} of the ATLAS detector and trigger based on GEANT4~\cite{Agostinelli:2002hh}. Finally, the 
simulated events are reconstructed and selected using the same analysis chain as is used for the collision data, with the same 
trigger and event selection criteria.

The flavour of jets is defined by matching jets to hadrons with $\pt > 5$\,GeV. The jet is considered a $b$-jet if a $b$-hadron is 
found within $\Delta R = \sqrt{\Delta\phi^2 + \Delta\eta^2} = 0.3$ of the jet axis; otherwise, if a $c$-hadron is found within the 
same distance the jet is labeled as a $c$-jet. All other jets are considered light\hyp{}flavour jets. 

The measured cross-sections are compared to NLO predictions derived using 
\powheg~\cite{Nason:2004rx,Frixione:2007vw,Alioli:2010xd,Frixione:2007nw} and 
\mcatnlo~\cite{Frixione:2002ik,Frixione:2003ei}, both using the MSTW 2008 NLO PDFs~\cite{Martin:2009iq} and a 
$b$-quark mass of 4.95\,GeV. To perform the parton\hyp{}showering, \powheg is interfaced to \pythia~6 and \mcatnlo to 
\herwig~6~\cite{Corcella:2000bw}. For Herwig, the AUET1~\cite{ATL-PHYS-PUB-2010-014} tune is used. In contrast to 
\pythia, \herwig uses an angular-ordered parton\hyp{}shower model and a cluster hadronisation model.


\section{Event and jet selection}

The events used in the lifetime\hyp{}based analysis are triggered by the L1 or HLT jet triggers, with the exception of the 
$20 < \pt < 40$\,GeV bin in the inclusive cross-section measurement where the MBTS trigger is used. The trigger efficiency for 
$b$-jets using these trigger selections is estimated to be above 97\% in all cases and typically close to 100\%. For the 
muon\hyp{}based cross-section measurement the combination of a jet and a muon trigger is required, which results in an 
efficiency ranging from about 35\% for jets with $\pt < 50$\,GeV to 65\% for jets with $\pt > 105$\,GeV. While this efficiency 
is lower, the different trigger prescale factors allocated a much higher rate to the jet-muon trigger than to the inclusive jet 
triggers for a similar jet \pt threshold.

Quality selections are applied to the reconstructed jets to ensure that they are not produced by poorly calibrated detector 
regions or noisy calorimeter cells~\cite{ATLAS-CONF-2010-038}. Additionally, the charged particle tracks contained in the jets 
are required to be of adequate quality for $b$-tagging~\cite{ATLAS-CONF-2010-042} and a good reconstructed primary vertex 
is required that contains at least 10 tracks with $\pt > 150$\,MeV. The combined efficiency of the reconstruction and the quality 
requirements is determined to be above 96\% for $b$-jets.

\begin{figure*}[th]
\begin{center}
\subfloat[Before $b$-tagging]{\includegraphics[width=0.495\textwidth]{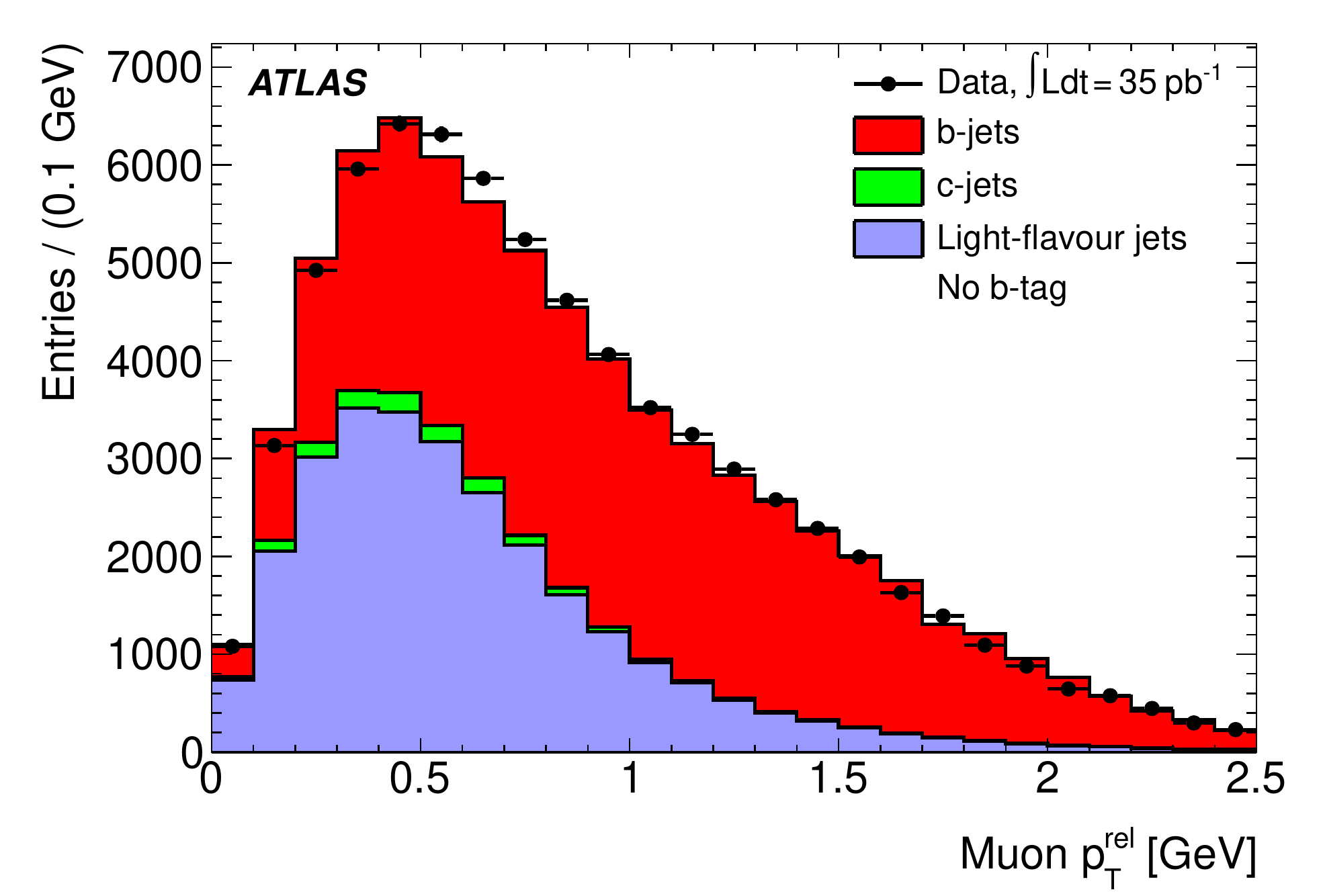}}
\subfloat[After $b$-tagging]{\includegraphics[width=0.495\textwidth]{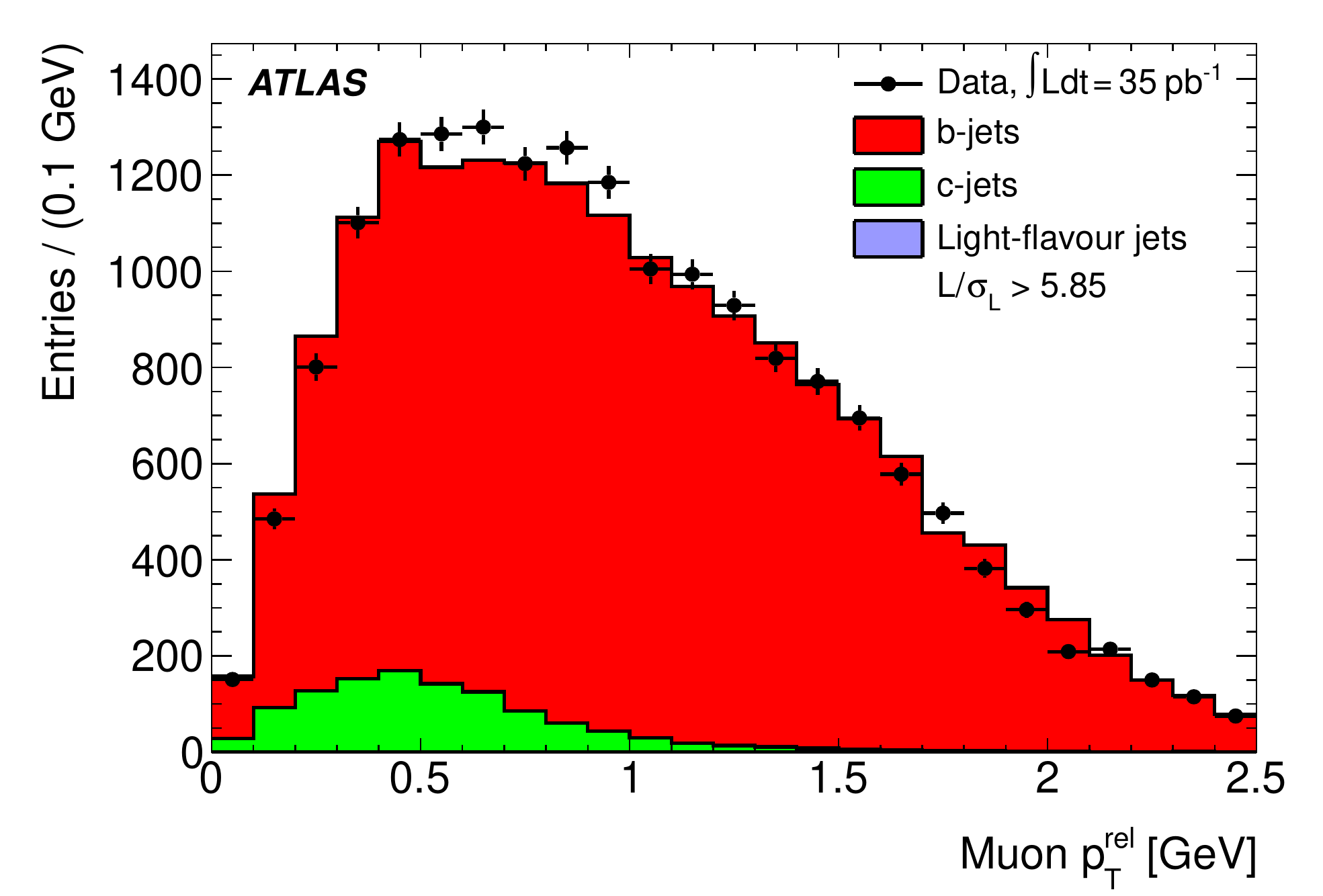}}
\caption{Examples of template fits to the measured \ptrel distribution, before and after applying the requirement of 
$L/\sigma_L > 5.85$. The error bars represent the data statistical errors. The differences between the data and the sum of the 
templates are covered by the systematic uncertainties on the template shapes.
\label{fig:ptrel_fits}}
\end{center}
\end{figure*}

The secondary vertex $b$-tagging algorithm used, SV0~\cite{ATLAS-CONF-2010-042}, aims at reconstructing the position of 
the displaced vertex from the charged decay products of long-lived particles in a jet. The SV0 algorithm reconstructs two-track 
vertices from tracks inside a cone of $\Delta R = 0.4$ around the jet axis that are significantly displaced from the primary vertex, 
based on the three\hyp{}dimensional impact parameter significance. Quality requirements are applied to the two-track vertices to 
reject vertices that are compatible with the primary vertex, are located at a radius consistent with one of the pixel detector layers 
or contain tracks that have an invariant mass consistent with a $K_S^0$ meson, a $\Lambda^0$ baryon or a photon conversion. 
A single secondary vertex is then fitted to all the tracks which contribute to any of the remaining two-track vertices in the jet.

The signed decay length significance of the secondary vertex, $L/\sigma_L$, is used to select a jet sample enriched in $b$-jets. 
The sign of the decay length is given by the sign of the projection of the decay length vector onto the jet axis. Jets with 
$L/\sigma_L > 5.85$ are referred to as $b$-tagged jets. The selection at 5.85 is chosen such that it produces a 50\% 
$b$-tagging efficiency for $b$-jets in simulated \ttbar events.

\subsection{$b$-tagging efficiency}

The efficiency of the chosen selection on $L/\sigma_L$ is estimated with a data-driven method that uses jets containing a 
muon. The number of $b$-jets before and after $b$-tagging can be obtained using the variable \ptrel, which is defined as the 
momentum of the muon transverse to the combined muon plus jet axis. Muons originating from $b$-hadron decays have a 
harder \ptrel spectrum than muons in $c$- and light\hyp{}flavour jets. Templates of the \ptrel shape are constructed for each 
jet flavour separately. The templates for $b$- and $c$-jets are extracted from Monte Carlo simulation, while the 
light\hyp{}flavour template is obtained from a light-jet enriched data sample. These are then fitted~\cite{Barlow:1993dm} to 
the measured \ptrel spectrum of muons in jets to obtain the fraction of $b$-jets before and after requiring a $b$-tag. The fit 
determines the relative contributions of the $b$-, $c$- and light\hyp{}flavour templates such that their sum best describes the 
shape of the \ptrel distribution in data. Having obtained the flavour composition of jets containing muons from the \ptrel fits, 
the $b$-tagging efficiency is defined as
\begin{displaymath}
e_b^{\rm data} =\frac{f_b^{\rm tag}\cdot N^{\rm tag}}{f_b\cdot N}\cdot C \,,
\end{displaymath}
where $f_b$ and $f_b^{\rm tag}$ are the fractions of $b$-jets before and after $b$-tagging is applied, and $N$ and 
$N^{\rm tag}$ are the total number of jets in those two samples. The factor $C$ corrects the efficiency for biases introduced 
by differences between data and simulation in the modelling of the $b$-hadron direction and by heavy\hyp{}flavour 
contamination of the \ptrel template for light\hyp{}flavour jets. The magnitude of these corrections is typically a few percent. 
Examples of fits to the \ptrel distribution before and after the $L/\sigma_L$ requirement are shown in Fig.~\ref{fig:ptrel_fits}.

The \ptrel method can be used to determine the $b$-tagging efficiency for $b$-jets containing $b$-hadrons that decay
semileptonically. Studies have been performed to show that this determination can be extended to all $b$-jets and a 
systematic uncertainty due to this generalization is assigned to the $b$-tagging efficiency for all $b$-jets. A detailed account of 
the systematic uncertainties in the $b$-tagging efficiency calibration is given in Ref.~\cite{ATLAS-CONF-2011-089}.

\begin{figure*}[th]
\begin{center}
\subfloat[Inclusive measurement]{\includegraphics[width=0.495\textwidth]{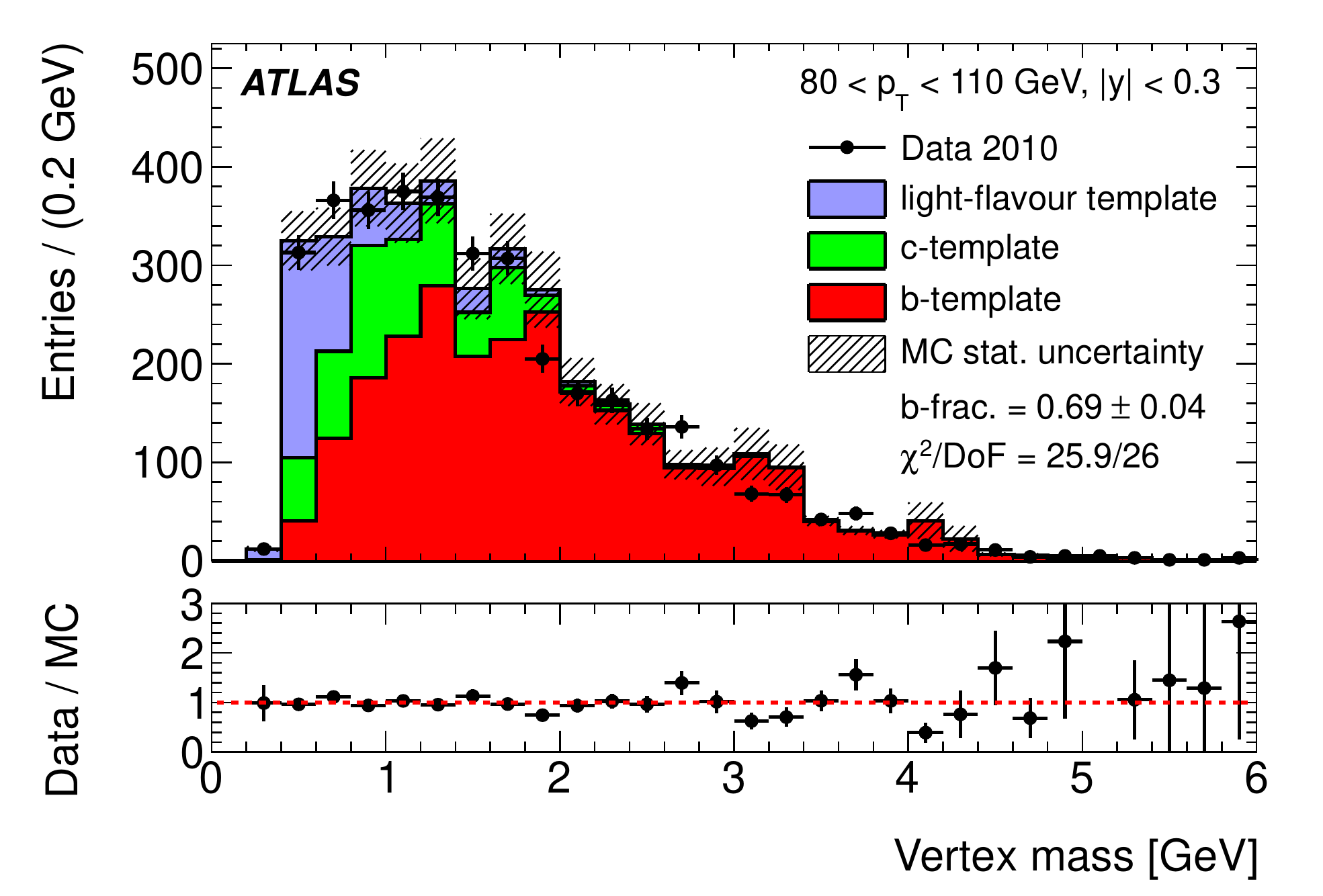}}
\subfloat[Dijet measurement]{\includegraphics[width=0.495\textwidth]{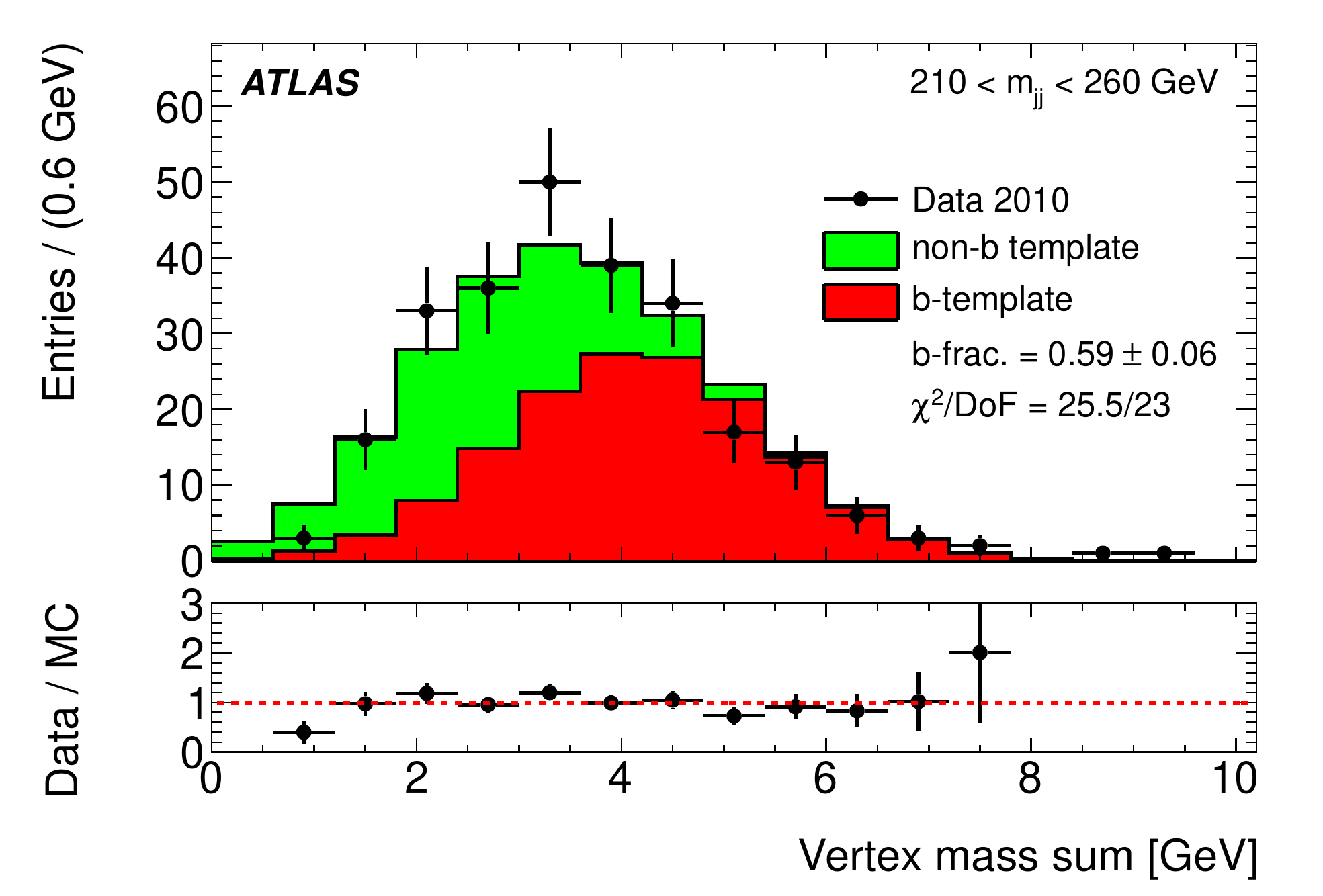}}
\caption{Examples of purity fits in the inclusive and dijet measurements. The error shown for the $b$-fraction is the uncertainty 
on the fit parameter. For the inclusive measurement the statistical uncertainty on the sum of the templates, indicated by the 
shaded area, is taken into account in the fit. In the dijet measurement the templates are parameterized, the uncertainty on the 
parameterization is taken into account as a systematic uncertainty and not shown here.}
\label{fig:fits}
\end{center}
\end{figure*}

The discriminating power of the \ptrel method decreases with increasing jet \pt, hence this method can only provide a 
data-driven determination of the $b$-jet tagging efficiency for jet \pt values up to about 140\,GeV. For jets with 
$\pt > 140$\,GeV, the $b$-tagging efficiency is derived from simulation and multiplied by a correction factor of $0.88 \pm 0.18$ 
that accounts for the difference between data and simulation observed in the \pt range 90--140\,GeV. Comparisons between data 
and simulation as a function of jet \pt show that the simulation models the data equally well in all regions of jet \pt in which data 
measurements are available, so the above extrapolation is well motivated. Moreover, detailed comparisons between data and 
simulation as a function of jet \pt, in terms of the quantities that affect $b$-tagging, show that the effect of any mismodelling of 
the $b$-tagging performance at higher jet \pt values is within the systematic uncertainties assigned to the $b$-tagging efficiency. 
The efficiency after applying the requirement of $L/\sigma_L > 5.85$ ranges from 20\% for $b$-jets of $\pt < 40$\,GeV and 
$|y| > 1.2$ to 55\% for central $b$-jets with \pt of about 100\,GeV.

The \ptrel distribution can also be used as a discriminant variable to measure the inclusive $b$-jet cross-section directly. While 
this method is statistically limited and cannot be used beyond 140\,GeV, as mentioned above, it does provide a useful 
cross-check for the lifetime\hyp{}based measurement. Many of the systematic uncertainties are different and the sample of 
jets used is statistically largely independent from that used in the lifetime\hyp{}based measurement. The muon-based 
cross-section measurement is described in Section~\ref{sec:results}.

\subsection{$b$-jet purity}

In the lifetime\hyp{}based measurement, the fraction of $b$-jets in the $b$-tagged sample of jets, referred to as the purity of 
the sample, is determined by performing a template fit to the secondary vertex mass distribution. The templates for $b$-, $c$- 
and light\hyp{}flavour jets are extracted from Monte Carlo simulation. The average invariant mass of a secondary vertex 
increases when going from light\hyp{}flavour jets via $c$-jets to $b$-jets, making it possible to separate the flavours by 
determining the relative fractions of the templates that best describe the vertex mass distribution in data.

For the inclusive cross-section measurement, the number of $b$-, $c$- and light\hyp{}flavour jets is fitted by maximizing a 
binned likelihood function that takes into account the statistical uncertainties in both the data and the templates. The fit is 
performed for each \pt and \y region separately, in vertex mass bins of 200\,MeV.

In the dijet cross-section measurement, the fraction of $b$-jet pairs is determined from a template fit to the sum of the vertex 
masses of the two $b$-tagged jets. This fit uses two templates: the $b$-template, where both jets are matched to a 
$b$-hadron in simulation; and a non-$b$ template, where at least one of the two jets is a $c$- or light\hyp{}flavour jet. In 
order to reduce the effect of the limited statistics in simulation, a parameterization is used to smooth the templates. The fit is 
performed for each kinematic region separately. Typical fit results in the inclusive and dijet measurements are shown in 
Fig.~\ref{fig:fits}.


\section{Results}\label{sec:results}

All the measured cross-sections are corrected for experimental effects using a bin-by-bin correction, so as to represent 
particle\hyp{}level cross-sections of jets containing $b$-hadrons. The correction is obtained from \pythia simulated dijet events by 
calculating the cross-sections for both particle\hyp{}level $b$-jets (including muons and neutrinos) and reconstructed $b$-jets. 
The correction factors are derived bin-by-bin in each distribution by taking the ratio of the two cross-sections. 

\subsection{Systematic uncertainties}

The dominant systematic uncertainties in both the inclusive and the dijet cross-section measurements come from the $b$-jet 
energy scale calibration, and the determination of the $b$-tagging efficiency and purity. The systematic uncertainties, including
those on the muon\hyp{}based measurement which will be discussed in Section~\ref{sec:ptrel}, are summarized in 
Table~\ref{tab:systematics}.

\begin{table}[ht]
\caption{Summary of the most important systematic uncertainties on the lifetime\hyp{}based inclusive $b$-jet and \bbbar-dijet 
cross-section, and on the muon\hyp{}based cross-section measurement.}
\label{tab:systematics}
\begin{tabular}{llll}
\hline\noalign{\smallskip}
Syst.\ uncertainty & Inclusive $b$-jet & \bbbar-dijet & Muon-based  \\
\noalign{\smallskip}\hline\noalign{\smallskip}
Jet energy scale & 10--20\% & 10--20\% & 15--20\% \\
$b$-tagging efficiency & 5--20\% & 30--50\% & -- \\
$b$-jet purity fit & 3--8\% & 20--30\% & 8--18\% \\
Luminosity & 3.4\% & 3.4\% & 3.4\% \\
Other sources & 2\% & 2\% & 3\% \\
\noalign{\smallskip}\hline
\end{tabular}
\end{table}

Jets are calibrated to the hadronic scale using the inclusive jet energy scale 
calibration~\cite{Aad:2010wv,ATLAS-CONF-2011-032}, which is based on \pt- and $\eta$-dependent correction factors 
derived from Monte Carlo simulation and validated with test beam measurements. The uncertainty on this jet energy scale 
varies between 2\% and 6\% depending on the jet \pt and rapidity region.

For heavy-flavour jets, two studies were performed to estimate additional contributions to the jet energy scale uncertainty that 
account for flavour-dependent systematic uncertainties. Firstly, the uncertainty on the calorimeter response for $b$-jets due to 
their different particle composition has been evaluated using single hadron response studies~\cite{ATLAS-CONF-2011-028}. This 
method compares the relative response of $b$-tagged jets in \ttbar events with that of inclusive jets in QCD dijet events. For jets 
within $|\eta| < 0.8$ and $20 < \pt < 250$\,GeV, this difference is found to be negligible ($<$0.5\%). Secondly, systematic 
uncertainties for $b$-jets were studied in Monte Carlo simulation by comparing particle\hyp{}level jets to reconstructed jets. The 
variations that were studied include the modelling of fragmentation, hadronisation, parton\hyp{}showers and the underlying 
event, but also variations in soft-physics tunes and the effects of the uncertainty on the material description. The $b$-jet energy 
scale uncertainty obtained using these two methods is validated in data by comparing the total transverse momentum of the 
calorimeter jet to that of the charged particle tracks associated to it~\cite{ATLAS-CONF-2011-067}. 

It is found that there is an additional 2.5\% uncertainty on the $b$-jet energy scale with respect to the uncertainty on the energy 
scale of inclusive jets. This extra uncertainty is added in quadrature. When propagated to the cross-section measurements, this 
leads to an uncertainty of 10\% to 20\%, depending on the kinematic region.

The most important contributions to the systematic uncertainty on the $b$-tagging efficiency originate from the modelling of 
muons in jets in the simulation, the generalization of the efficiency from $b$-jets with muons to inclusive $b$-jets, and the 
limited statistics of the templates used for the \ptrel fits. More details about the $b$-tagging efficiency uncertainty can be 
found in Ref.~\cite{ATLAS-CONF-2011-089}. The resulting uncertainty on the cross-sections amounts to between 5\% and 
20\% for the inclusive $b$-jet cross-section, and between 30\% and 50\% for the dijet cross-section.

The systematic uncertainties from the purity fits account for the observed differences between jets in collision data and those in 
the Monte Carlo simulation used to derive the templates. The uncertainty is derived from studies of the secondary vertex 
mass distribution in light-jet enriched samples and $b$-jet enriched samples. The light-jet enriched sample is obtained by 
selecting jets with a negative decay length. For the $b$-jet enriched samples, two methods are used: the first requires another 
$b$-tagged jet to be present in the event, while the second selects secondary vertices with high track multiplicities. The 
observed differences in the secondary vertex mass distribution are then used to correct the template shapes and re-evaluate 
the fits. The difference in the cross-section is found to be between 3\% and 8\% and this is assigned as the purity fit 
systematic uncertainty. For the $b\overline{b}$-dijet cross-section, the most important contribution to the systematic 
uncertainty on the $b\overline b$-fraction is due to the limited template statistics. The effect of the statistical uncertainty of the 
templates is estimated by varying the shape parameters of the parameterized templates within their uncertainties and 
re-evaluating the cross-section. The resulting uncertainty is between 20\% and 30\%.

The systematic uncertainty on the luminosity determination is 3.4\%~\cite{ATLAS-CONF-2011-011}. The remaining sources of 
systematic uncertainty, such as the effect of possible differences in the cross-section shapes between data and simulation on 
the bin-by-bin corrections, differences in the jet energy resolution between data and simulation, the trigger efficiency and the jet 
selection efficiency, lead to a combined systematic uncertainty of about 2\%.

The effect of different shower and hadronisation models is included in the jet energy scale uncertainty. The impact of changing 
the shape of the \pt distribution on the bin-by-bin corrections was found to be much less than 1\%. Using \herwig instead of 
\pythia to derive the correction factors gives statistically consistent results.

\subsection{Muon\hyp{}based $b$-jet cross-section}\label{sec:ptrel}

The \ptrel method, used for calibrating the $b$-tagging efficiency, is also used to obtain an independent measurement of the 
inclusive $b$-jet cross-section in the range $30 < \pt < 140$\,GeV. This measurement uses jets containing a muon of 
$\pt > 4$\,GeV within a cone of $\Delta R = 0.4$ from the jet axis. The flavour composition of this sample is extracted from a 
template fit to the muon \ptrel distribution. The templates for $b$- and $c$-jets are obtained from Monte Carlo simulation. Two 
data-driven techniques are employed to extract the shape of the muon \ptrel in light\hyp{}flavour jets. The first takes the shape 
from jets with negative decay length in data, which is then corrected for $b$-jet contamination using simulation. The second 
method uses inclusive jets without a muon; the template is then obtained by converting each track inside the jet into a muon 
and weighting the resulting \ptrel by a probability to simulate hadron decays in flight. The $b$-jet fraction is evaluated using 
both methods, taking the average as the central value and assigning the difference between them as a systematic uncertainty.

The dominant sources of systematic uncertainties in this measurement are the $b$-jet energy scale (15--20\%) and the purity 
fits (8--18\%). Contributions to the purity fit systematics include limited template statistics and uncertainties in the modelling of 
semileptonic $b$-hadron decays and $b$-fragmentation. The first modelling error is estimated by varying the muon 
momentum distribution in the rest frame of the $b$-hadron between that measured by DELPHI~\cite{Abdallah:2005cx} and that 
measured by BABAR~\cite{Aubert:2004td}. The second is measured by varying the fraction of the $b$-jet energy carried by the 
$b$-hadron by $\pm 5$\% and rederiving the $b$-jet templates in the simulation. Apart from the $b$-jet energy scale, the 
systematic uncertainties are to a large extent specific to the muon\hyp{}based measurement. This makes the comparison 
with the lifetime\hyp{}based cross-section measurement a useful cross-check.

\subsection{Cross-section results and discussion}

\begin{figure}[th]
\begin{center}
\includegraphics[width=0.485\textwidth]{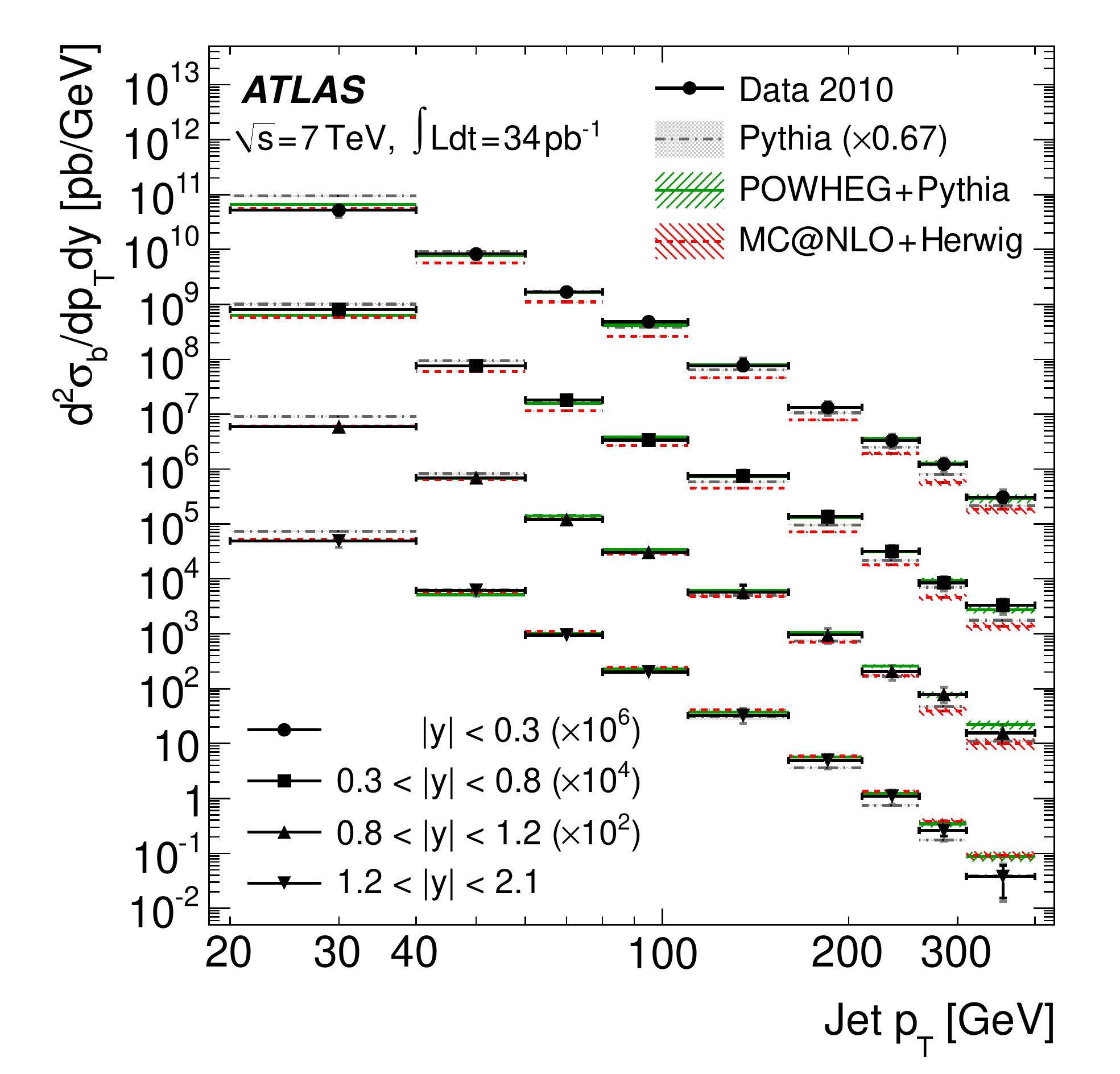}
\caption{Inclusive double-differential $b$-jet cross-section as a function of \pt for the different rapidity ranges. The data are 
compared to the predictions of \pythia, \powheg and \mcatnlo. The leading-order \pythia prediction is scaled $(\times 0.67)$ to 
the measured integrated cross-section.}
\label{fig:xsec}
\end{center}
\end{figure}

\begin{figure}[th]
\begin{center}
\includegraphics[width=0.485\textwidth]{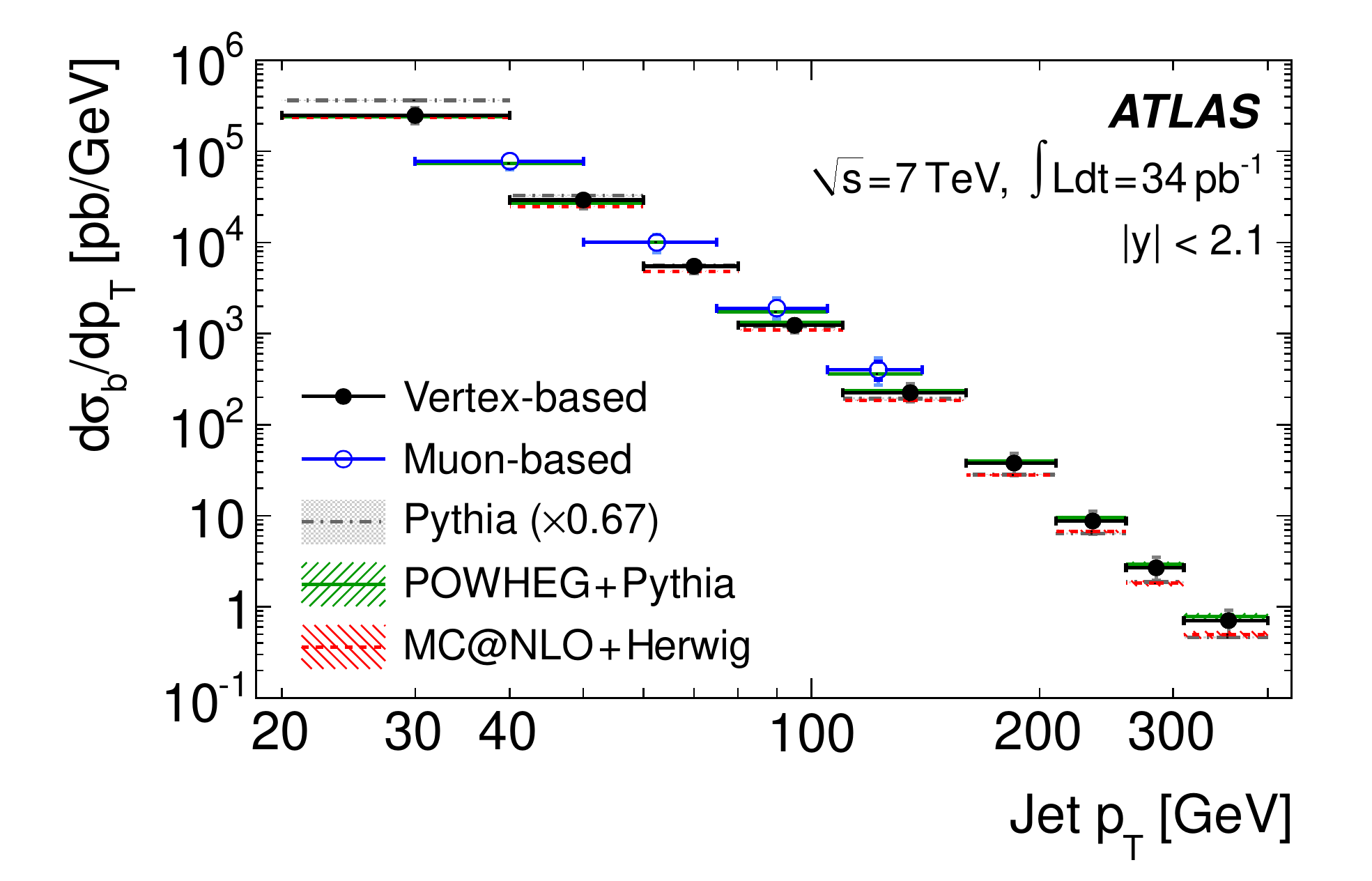}
\caption{Differential $b$-jet cross-section as a function of \pt for $b$-jets with $|y| < 2.1$. The data are compared to the 
predictions of \pythia, \powheg and \mcatnlo. In the region $30 < \pt < 140$\,GeV the muon-based cross-section measurement 
is also shown. For the muon-based measurement only the \powheg prediction is shown.}
\label{fig:intxsec}
\end{center}
\end{figure}

The double-differential inclusive $b$-jet cross-section is shown in Fig.~\ref{fig:xsec} as a function of jet \pt in four different 
rapidity regions. Figure~\ref{fig:intxsec} shows the single differential cross-section as a function of \pt, integrated over the 
entire rapidity range of $|y| < 2.1$. In the \pt range where the lifetime\hyp{}based and the muon\hyp{}based measurements 
overlap, both results are shown. The data are compared to NLO predictions derived with \powheg and \mcatnlo. In addition, 
the data are compared to the \pythia prediction. \pythia, as a leading-logarithmic parton\hyp{}shower generator, is not 
expected to predict the correct normalization. The \pythia prediction is scaled by a factor $\times 0.67$ in order to match the 
measured integrated cross-section, allowing a comparison of the cross-section shapes. All three calculations describe the 
general features of the cross-section reasonably well.

\begin{figure*}[tp]
\begin{center}
\subfloat[\powheg]{\label{fig:xsecratio_powheg}\includegraphics[width=0.495\textwidth]{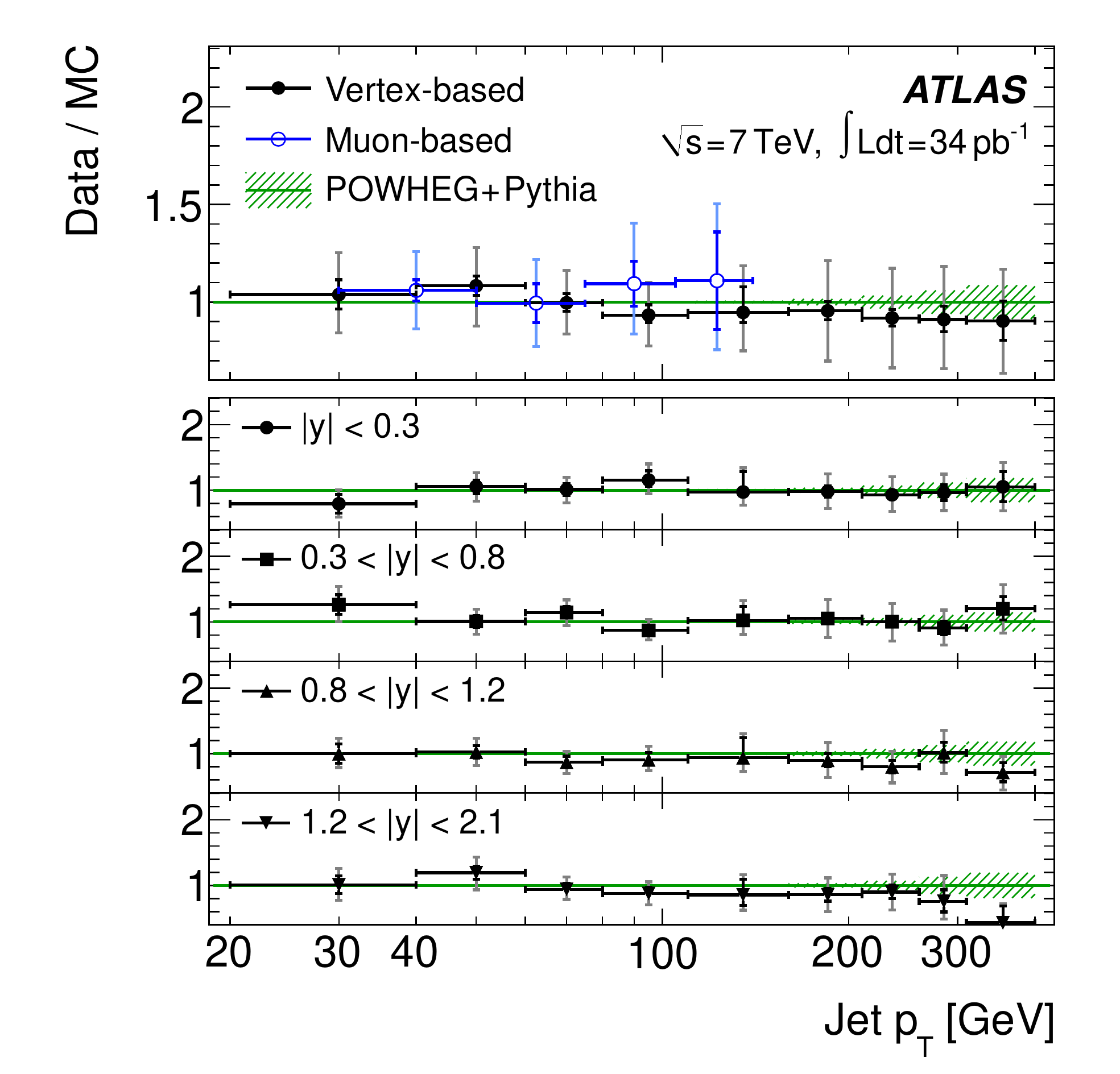}}
\subfloat[\mcatnlo]{\label{fig:xsecratio_mcatnlo}\includegraphics[width=0.495\textwidth]{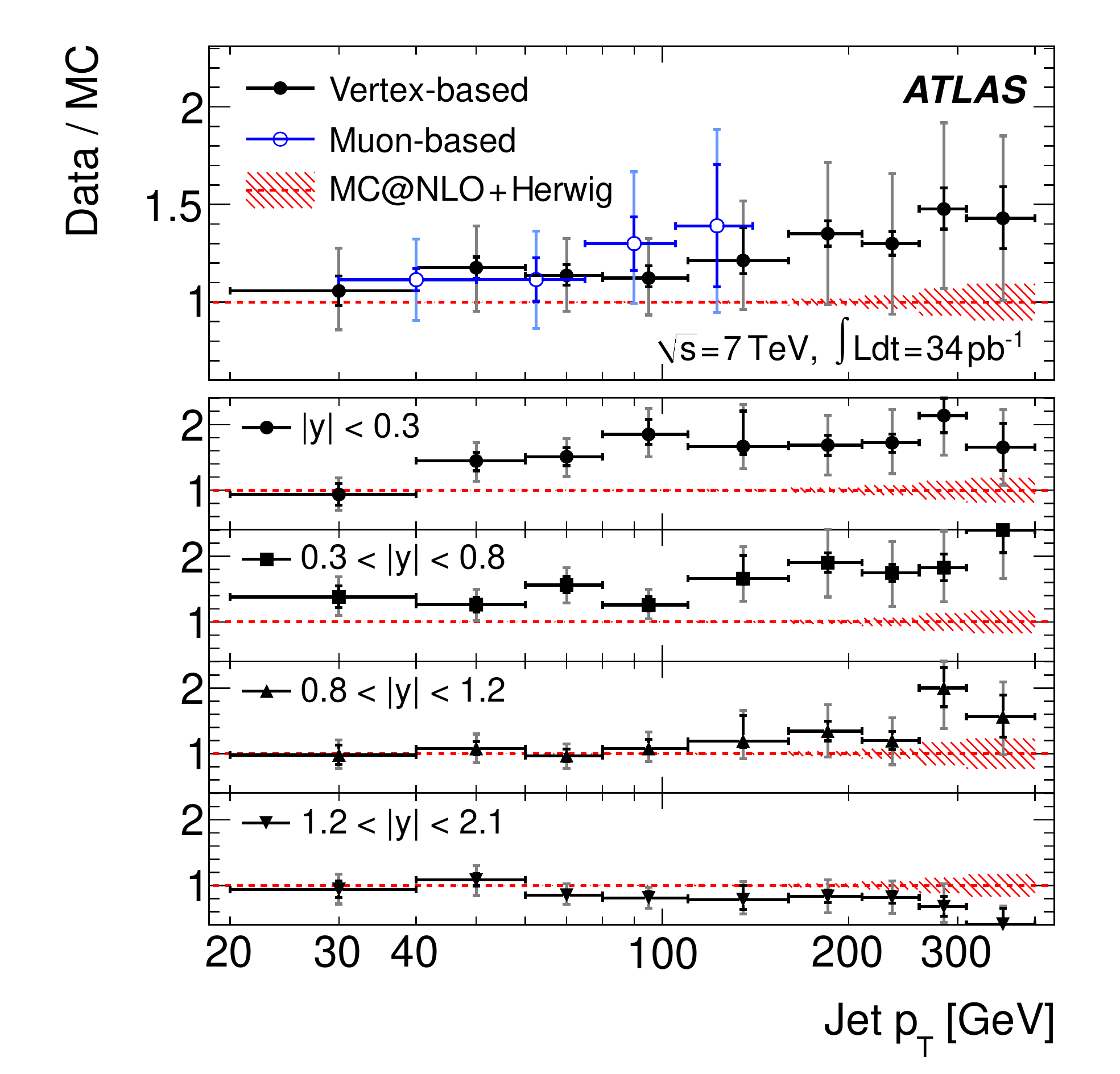}}
\caption{Ratio of the measured cross-sections to the theory predictions of \powheg and \mcatnlo. In the region where the 
lifetime\hyp{}based measurement overlaps with the muon \ptrel measurement both results are shown. The top plot shows the 
full rapidity acceptance, while the four smaller plots show the comparison for each of the rapidity ranges separately. The data 
points show both the statistical uncertainty (dark colour) and the combination of the statistical and systematic uncertainty (light 
colour). The shaded regions around the theoretical predictions reflect the statistical uncertainty only. Systematic uncertainties in 
the NLO predictions are discussed in the text.}
\label{fig:xsecratio}
\end{center}
\end{figure*}

To allow for a better comparison between the data and the NLO predictions, Fig.~\ref{fig:xsecratio} shows the ratio of the 
measured cross-section to the NLO theory predictions for $|y| < 2.1$ (top) and for each rapidity region separately. The plot for 
the full rapidity acceptance also allows a direct comparison between the lifetime\hyp{}based and the muon\hyp{}based 
cross-section measurements in the overlapping \pt range, indicating a good agreement between the two measurements. Good 
agreement is also observed between the measured cross-section and the NLO predictions obtained using \powhegpythia in all 
rapidity regions. \mcatnloherwig, however, predicts a significantly different behaviour of the double-differential cross section, as 
shown in Fig.~\ref{fig:xsecratio_mcatnlo}. When the cross-section is integrated over the full rapidity acceptance this effect 
averages out somewhat and \mcatnloherwig shows better agreement with data. It has been checked that the qualitative 
behaviour remains the same when \powheg is interfaced to \herwig instead of \pythia, implying that the observed rapidity 
dependence in \mcatnloherwig is not resulting from the parton\hyp{}shower Monte Carlo program. On the other hand, 
\powhegherwig appears to predict a cross-section that is consistently lower than the \powhegpythia prediction. This would
suggest that the deficit of \mcatnloherwig compared to the data in Fig.~\ref{fig:xsecratio_mcatnlo}, may be partly due to the 
\herwig parton\hyp{}showering.

Comparison to the inclusive (all-flavour) jet cross-section measurement~\cite{Aad:2010wv}, shows that the fraction of 
jets containing a $b$-hadron is approximately 5\% in the kinematic region where the two measurements overlap, 
$60 < \pt < 400$\,GeV and $|y| < 2.1$.

\begin{figure}[thp]
\begin{center}
\includegraphics[width=0.485\textwidth]{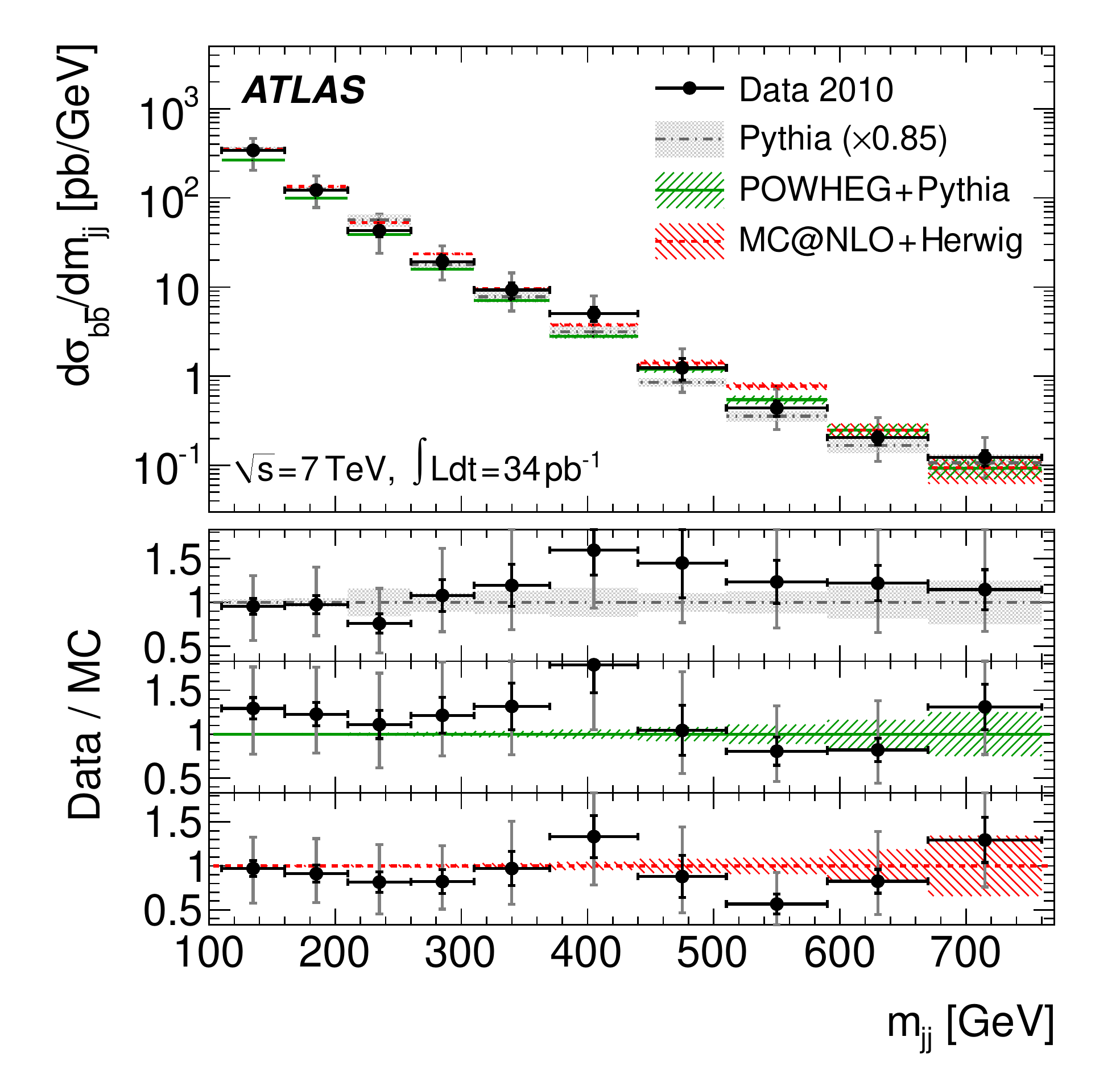}
\caption{The \bbbar-dijet cross-section as a function of dijet invariant mass for $b$-jets with $\pt>40$\,GeV and $|y|<2.1$. 
The data are compared to the MC predictions of \pythia, \powheg and \mcatnlo. The leading-order \pythia prediction is scaled 
to the measured integrated cross-section. The shaded regions around the MC predictions reflect the statistical uncertainty 
only.}
\label{fig:dijet-mass}
\end{center}
\end{figure}

The \bbbar-dijet cross-section is shown as a function of dijet mass in Fig.~\ref{fig:dijet-mass}. It should be noted that nearby 
\bbbar-pairs, as expected for example from gluon splitting, are generally not resolved as separate jets. Also, since the 
measurement refers to the leading and sub-leading jet in the event, the contribution from gluon splitting is expected to be 
small. The \bbbar-dijet cross-section is compared to \pythia and the NLO predictions obtained using \powheg and \mcatnlo. 
The \pythia prediction is again normalized to the measured integrated cross-section, here using a factor of $\times0.85$. The 
\pythia normalization is not expected to be the same as that used in the inclusive cross-section, given the different event selection 
used. All theory predictions show good agreement with the measured cross-section.

\begin{figure}[thp]
\begin{center}
\includegraphics[width=0.485\textwidth]{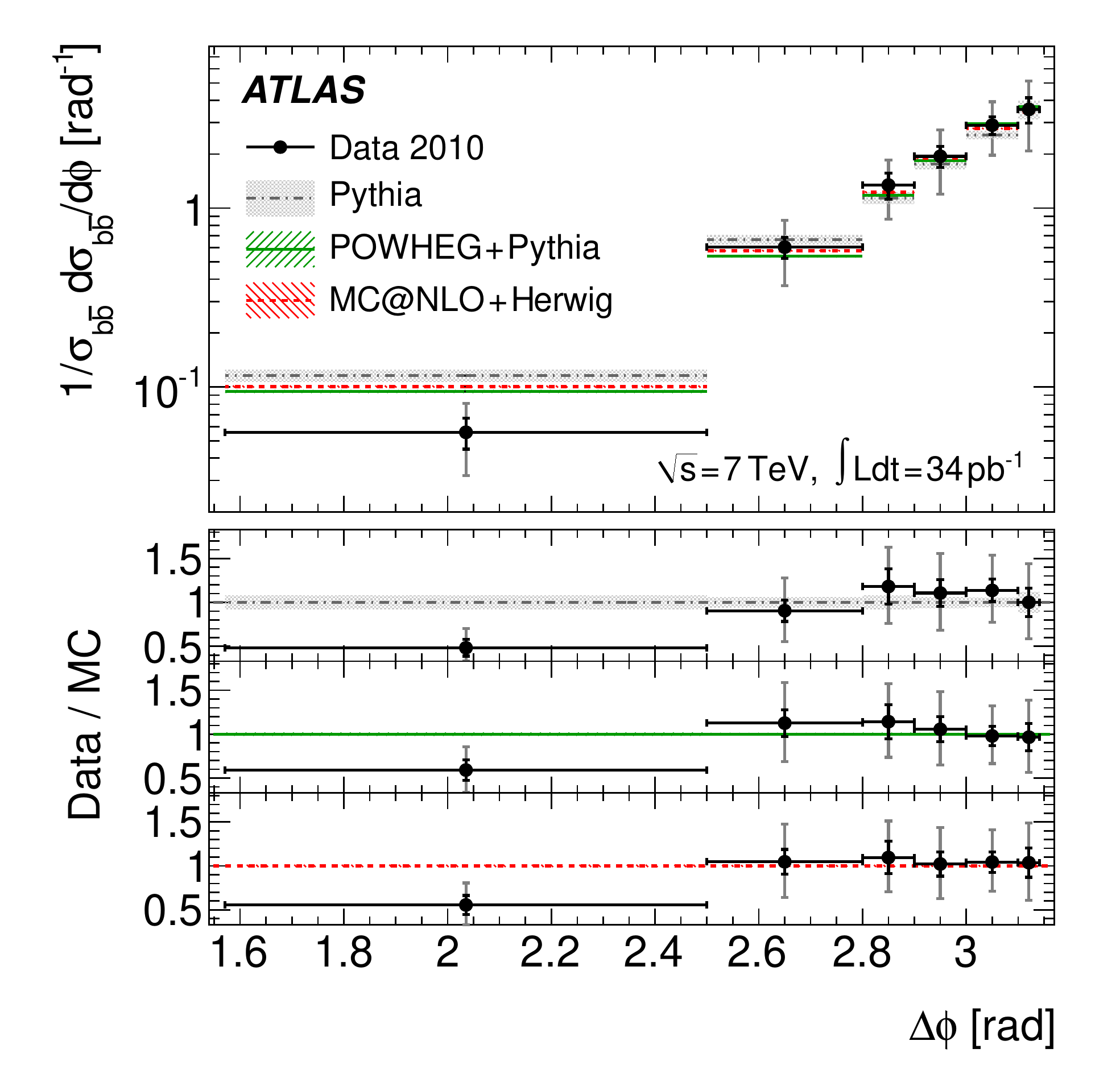}
\caption{The \bbbar-dijet cross-section as a function of the azimuthal angle difference between the two jets for $b$-jets with 
$\pt>40$\,GeV, $|y|<2.1$ and a dijet invariant mass of $\mjj > 110$\,GeV. The data are compared to the theory predictions 
of \pythia, \powheg and \mcatnlo. The shaded regions around the MC predictions reflect the statistical uncertainty only.}
\label{fig:dijet-phi}
\end{center}
\end{figure}

Figure~\ref{fig:dijet-phi} shows the fractional \bbbar-dijet cross-section as a function of the azimuthal angle between the 
two jets, $\Delta\phi$. The dijets selected in this measurement show a pronounced back-to-back configuration in the 
transverse plane that is generally well reproduced by QCD generators.

\begin{figure}[thp]
\begin{center}
\subfloat[$110 < m_{\rm jj} < 370$\,GeV]{\includegraphics[width=0.485\textwidth]{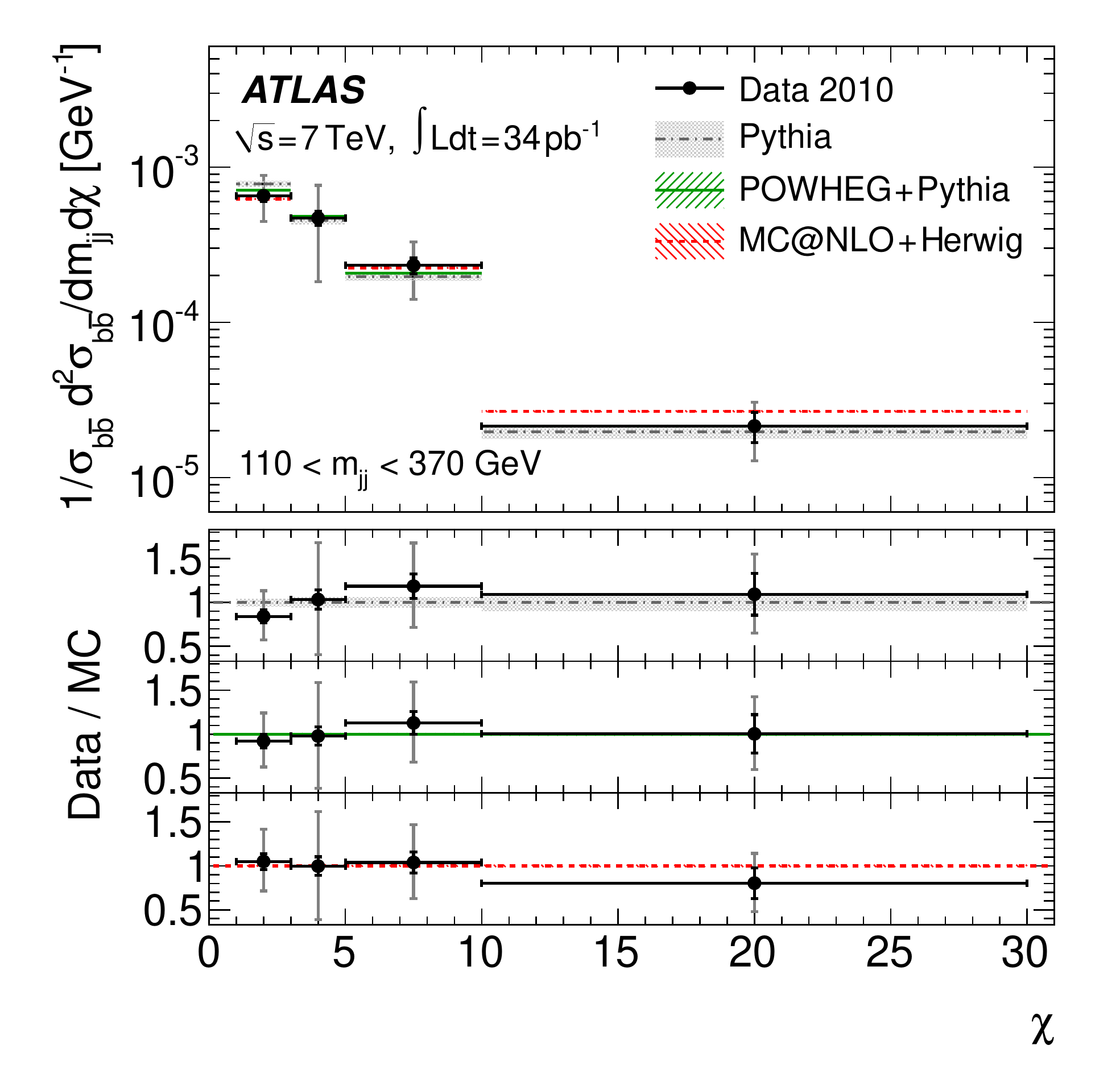}} \\
\subfloat[$370 < m_{\rm jj} < 850$\,GeV]{\includegraphics[width=0.485\textwidth]{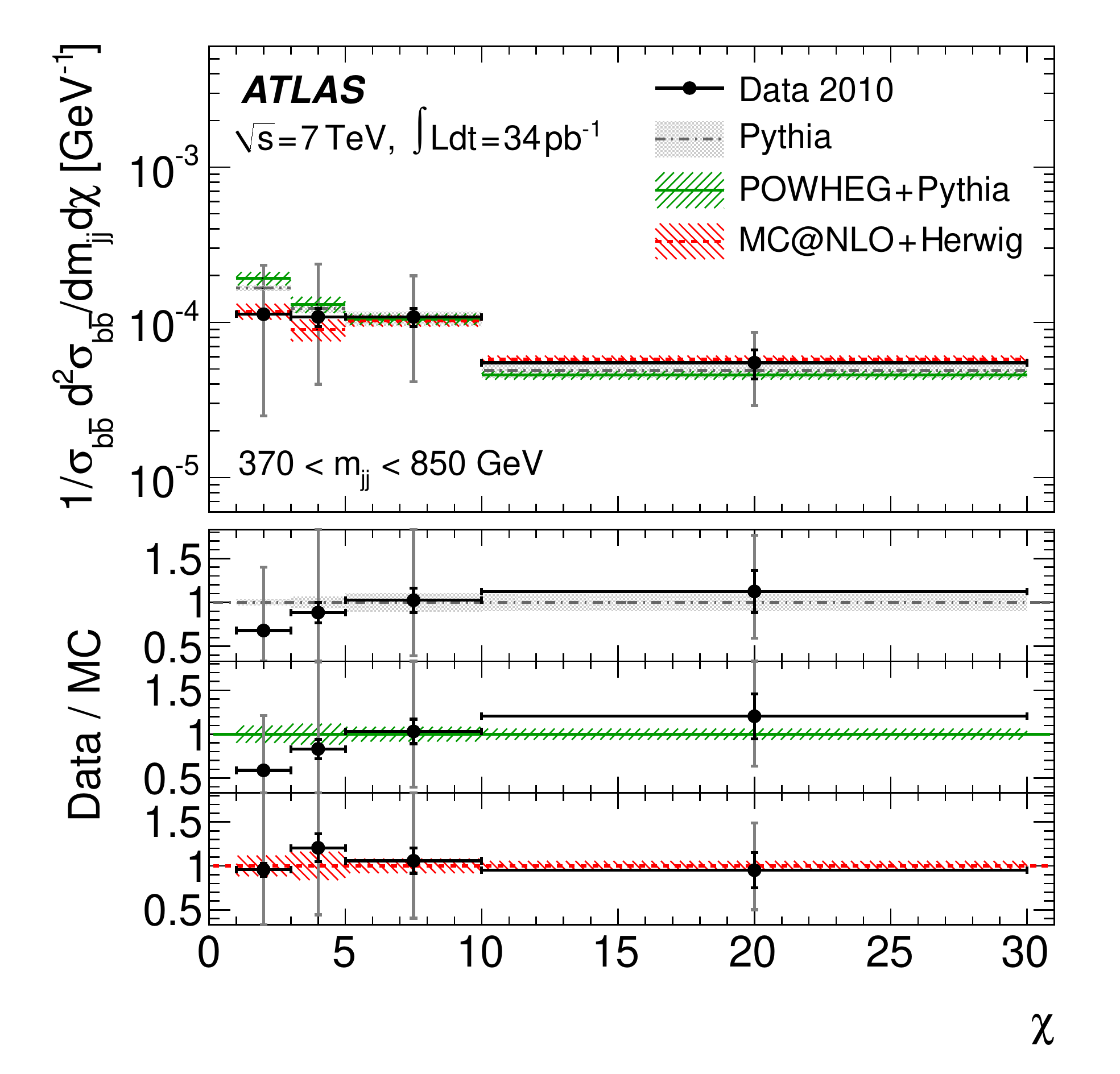}}
\caption{The \bbbar-dijet cross-section as a function of $\chi$ for $b$-jets with $\pt>40$\,GeV, $|y|<2.1$ and 
$|y_{\rm boost}| = \frac{1}{2}|y_1+y_2| < 1.1$, for two dijet invariant mass ranges. The data are compared to the theory 
predictions of \pythia, \powheg and \mcatnlo. The shaded regions around the MC predictions reflect the statistical uncertainty 
only.}
\label{fig:dijet-chi}
\end{center}
\end{figure}

The $b\overline b$-dijet cross-section as a function of the angular variable $\chi$ is shown in Fig.~\ref{fig:dijet-chi} for dijets 
with $|y_{\rm boost}| < 1.1$. The $\chi$ distribution is well reproduced by the theoretical calculations. The distribution flattens 
for large invariant mass values.

In the NLO calculations, the renormalization and factorization scales are set equal to the transverse energy of the hardest parton: 
$Q^2 = E_{\rm T}^2 = m_b^2 + \pt^2$. To estimate the potential impact of higher order terms not included in the NLO 
calculation on the theory predictions, the renormalization scale is varied from half to twice its default value. Similarly, to estimate 
the impact of the choice of the scale where the PDF evolution is separated from the matrix element, the factorization scale is 
varied up and down by a factor of two. The effect of each of these variations on the NLO cross-section prediction is estimated 
using \powheg and found to be approximately 20\% for all kinematic regions. Finally, the uncertainty on the PDFs is estimated by 
deriving the NLO predictions using the NNPDF~\cite{Ball:2010de} and CTEQ~6.6~\cite{Nadolsky:2008zw} PDFs, resulting in a 
difference of approximately 10\% for all kinematic regions.


\section{Conclusions}

The inclusive $b$-jet and \bbbar-dijet production cross-sections have been measured in proton--proton collisions at a 
centre-of-mass energy of 7\,TeV, using data with an integrated luminosity of 34\,pb$^{-1}$ recorded by the ATLAS detector.

The inclusive $b$-jet cross-section was measured as a function of jet \pt in the range $20 < \pt < 400$\,GeV and rapidity in 
the range $|y| < 2.1$. The \bbbar-dijet cross-section was measured as a function of dijet invariant mass in the range 
$110 < \mjj < 760$\,GeV, as a function of the azimuthal angle difference and of the angular variable $\chi$. The 
measurements are dominated by systematic uncertainties, mainly coming from the $b$-jet energy scale and the determination 
of the $b$-jet tagging efficiency and purity. The measured cross-sections have been compared to next-to-leading order QCD 
predictions derived using \powheg interfaced to \pythia and \mcatnlo interfaced to \herwig. 

The inclusive cross-section measured over $|y| < 2.1$ for $b$-jets identified by the presence of a secondary vertex is compared 
to a largely independent cross-section measurement that uses muon-based $b$-tagging in the range $30 < \pt < 140$\,GeV. The 
two measurements show good agreement.

The inclusive $b$-jet cross-section is found to be in good agreement with the \powhegpythia prediction over the full kinematic 
range. \mcatnloherwig, however, predicts a significantly different behaviour of the double-differential cross section that is not 
observed in the data. The normalized leading-order \pythia prediction shows broad agreement with the measured cross-section.

\powhegpythia and \mcatnloherwig show good agreement with the measured \bbbar-dijet cross-sections, as does the 
normalized leading-order \pythia generator.


\section*{Acknowledgements}

{\tolerance1500

We thank CERN for the very successful operation of the LHC, as well as the
support staff from our institutions without whom ATLAS could not be
operated efficiently.

We acknowledge the support of ANPCyT, Argentina; YerPhI, Armenia; ARC,
Australia; BMWF, Austria; ANAS, Azerbaijan; SSTC, Belarus; CNPq and FAPESP,
Brazil; NSERC, NRC and CFI, Canada; CERN; CONICYT, Chile; CAS, MOST and
NSFC, China; COLCIENCIAS, Colombia; MSMT CR, MPO CR and VSC CR, Czech
Republic; DNRF, DNSRC and Lundbeck Foundation, Denmark; ARTEMIS, European
Union; IN2P3-CNRS, CEA-DSM/IRFU, France; GNAS, Georgia; BMBF, DFG, HGF, MPG
and AvH Foundation, Germany; GSRT, Greece; ISF, MINERVA, GIF, DIP and
Benoziyo Center, Israel; INFN, Italy; MEXT and JSPS, Japan; CNRST, Morocco;
FOM and NWO, Netherlands; RCN, Norway; MNiSW, Poland; GRICES and FCT,
Portugal; MERYS (MECTS), Romania; MES of Russia and ROSATOM, Russian
Federation; JINR; MSTD, Serbia; MSSR, Slovakia; ARRS and MVZT, Slovenia;
DST/NRF, South Africa; MICINN, Spain; SRC and Wallenberg Foundation,
Sweden; SER, SNSF and Cantons of Bern and Geneva, Switzerland; NSC, Taiwan;
TAEK, Turkey; STFC, the Royal Society and Leverhulme Trust, United Kingdom;
DOE and NSF, United States of America.

The crucial computing support from all WLCG partners is acknowledged
gratefully, in particular from CERN and the ATLAS Tier-1 facilities at
TRIUMF (Canada), NDGF (Denmark, Norway, Sweden), CC-IN2P3 (France),
KIT/GridKA (Germany), INFN-CNAF (Italy), NL-T1 (Netherlands), PIC (Spain),
ASGC (Taiwan), RAL (UK) and BNL (USA) and in the Tier-2 facilities
worldwide.

}


\bibliographystyle{atlasnote}
\bibliography{bjetpaper}


\appendix

\clearpage
\onecolumn
\section*{The ATLAS Collaboration}

\begin{flushleft}
G.~Aad$^{\rm 48}$,
B.~Abbott$^{\rm 111}$,
J.~Abdallah$^{\rm 11}$,
A.A.~Abdelalim$^{\rm 49}$,
A.~Abdesselam$^{\rm 118}$,
O.~Abdinov$^{\rm 10}$,
B.~Abi$^{\rm 112}$,
M.~Abolins$^{\rm 88}$,
H.~Abramowicz$^{\rm 153}$,
H.~Abreu$^{\rm 115}$
% ...
\bigskip

% ...
$^{10}$ Institute of Physics, Azerbaijan Academy of Sciences, Baku, Azerbaijan\\
$^{11}$ Institut de F\'isica d'Altes Energies and Departament de F\'isica de la Universitat Aut\`onoma  de Barcelona and ICREA, Barcelona, Spain\\
% ...
$^{48}$ Fakult\"{a}t f\"{u}r Mathematik und Physik, Albert-Ludwigs-Universit\"{a}t, Freiburg i.Br., Germany\\
$^{49}$ Section de Physique, Universit\'e de Gen\`eve, Geneva, Switzerland\\
% ...
$^{88}$ Department of Physics and Astronomy, Michigan State University, East Lansing MI, United States of America\\
% ...
$^{111}$ Homer L. Dodge Department of Physics and Astronomy, University of Oklahoma, Norman OK, United States of America\\
$^{112}$ Department of Physics, Oklahoma State University, Stillwater OK, United States of America\\
% ...
$^{115}$ LAL, Univ. Paris-Sud and CNRS/IN2P3, Orsay, France\\
% ...
$^{118}$ Department of Physics, Oxford University, Oxford, United Kingdom\\
% ...
$^{153}$ Raymond and Beverly Sackler School of Physics and Astronomy, Tel Aviv University, Tel Aviv, Israel\\
% ...
\end{flushleft}

\end{document}